# Angle-Resolved Photoemission Spectroscopy Studies of Curpate Superconductors

ZHI-XUN SHEN

*Department of Applied Physics, Physics and Stanford Synchrotron Radiation Laboratory, Stanford University, Stanford, CA 94305*

This paper summarizes experimental results presented at the international conference honoring Prof. C.N. Yang's 80[th] birthday. I show seven examples that illustrate how one can use angle-resolved photoemission spectroscopy to gain insights into the many-body physics responsible for the rich phase diagram of cuprate superconductors. I hope to give the reader a snapshot of the evolution of this experimental technique from a tool to study chemical bonds and band structure to an essential many-body spectroscopy for one of the most important physics problems of our time.

Complex phenomenon in solids is going to be a major theme of physics in the 21[st] century. As better controlled model systems, a sophisticated understanding on the universality and diversity of these solids may lead to great revelations well beyond themselves. With its rich phases and extremely high superconducting transition temperature, the cuprate superconductor is the most dramatic example of complex phenomena in solids and is thus the most challenging and important problem of the field over the last two decades. High-resolution angle-resolved photoemission spectroscopy (ARPES) has emerged as a leading tool to push the frontier of this important field of modern physics. It helped setting the intellectual agenda by testing new ideas, discovering surprises, and challenging orthodoxies. Indeed, quite a few ARPES papers are among the most cited physics papers for the periods surveyed by the Institute of Scientific Information [1-5]. There is little doubt that this technique is going to be at the focal point of the necessary debates leading to new paradigms of physics – those gone well beyond the Fermi liquid paradigm so dominated the solid state physics textbooks today. It is likely that what we have seen is just a tip of the iceberg. This paper highlights some of the progresses over the last decade and discusses the prospects for future development.

Improved resolution and carefully matched experiments have been the keys to turn this technique from a chemical analysis and band mapping tool into a sophisticated many-body spectroscopy. Fundamentally, the power of the technique stems from its directness and richness in information. Some physics of solids are already understood by their macroscopic and thermodynamic properties, but the truly deep insights often come from scattering experiments. As a special form of scattering experiments, ARPES provides what we need the most: the direction, the speed, and the scattering processes of valence electrons. With the extremely high angular and energy resolutions now achievable, the technique reveals the electronic structure with unprecedented precision and sophistication- information that forms the foundation for a comprehensive understanding of complex solids. Indeed, it merits what has sometimes been called: a microscope for where and how the electrons move. There is no other tool that can equally well visualize the energy-momentum phase space of the electrons - a world necessary to know but difficult to feel.

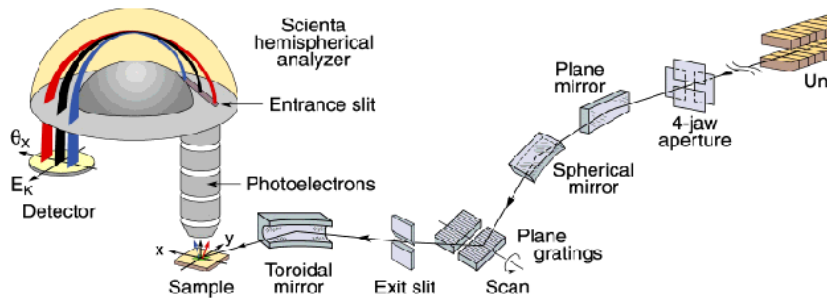

Figure 1. A typical ARPES set-up.

A typical ARPES experiment is carried out on freshly prepared sample surfaces in ultra-high vacuum as the short mean free path of the photoelectrons dictates that only the near surface electrons carry the inherent information without suffering scatterings. The surface sensitivity can be a problem for ARPES experiments. However, the two dimensional nature and the easy cleavage of the cuprate crystals minimizes the problem although special attention is still required in certain cases. Largely stimulated by the experimental need, the last fifteen years saw a dramatic improvement in energy and angular (thus momentum) resolution, from typical 200 meV and 2~4 degree to 2~10 meV and 0.1-0.3 degree respectively. This advance was made possible by the development of sophisticated spectrometers and advanced synchrotron beamlines, as illustrated in figure 1 for an undulator beamline with two-dimensional photoelectron detection scheme. This advance happened steadily over the last decade and at each stage carefully matched experiments were carried out to address the cuprate problem. This effort was aided by the remarkable progress in high-quality single crystal growth, a triumph in material physics that enabled a wide range of high quality experiments, including ARPES.

Under the sudden approximation, ARPES measures the single particle spectral function $A(\mathbf{k},\omega)$ weighted by the photo-ionization cross-section and the Fermi-Dirac function. In a more formal language, the spectral function is the imaginary part of the single particle Green's function $G(\mathbf{k},\omega)$. Because the Green's function can be calculated from microscopic many-body Hamiltonians, ARPES results often provide a direct test for theory without the need to make ensemble averages as is the case for most other experiments. Indeed, the directness and the microscopic nature of the measurement is the main source of power for these experiments. In the context of superconductors, APRES experiments measure a quantity similar to that measured routinely by tunneling experiments, the main technique for conventional superconductors (albeit ARPES provides additional $\mathbf{k}$-resolved information). In a non-interacting system, both ARPES and tunneling reduce to simple density of state information. We note that modern tunneling experiments provide $\mathbf{r}$-resolved spectra that can lead to $\mathbf{k}$-resolved information, but that is beyond the scope of this paper.

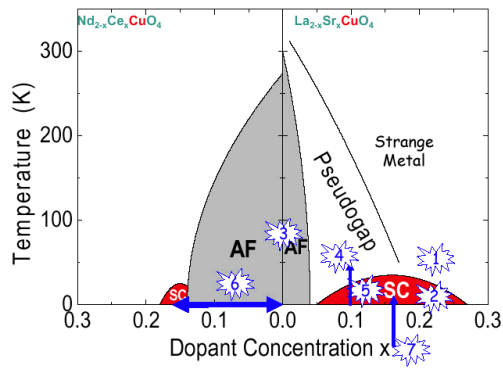

Figure 2. Phase diagram of p- and n-type cuprates. The numbers marked indicate the regions where examples of ARPES results will be discussed in this paper.

I will organize the discussion around the question on how one can use ARPES to gain more insight on the complex phase diagram of cuprate superconductors. Figure 2 presents the well-known phase diagram of cuprates: at low temperature the material is an antiferromagnetic insulator when undoped, and becomes a high-temperature superconductor at an appropriate doping level and eventually a non-superconducting metal at sufficiently high doping level. In the superconducting state, we call the doping range where $T_c$ has not yet reached maximum underdoped, where $T_c$ reaches maximum optimal, and where $T_c$ again decreases with doping the overdoped regimes, respectively. At elevated temperatures, the so-called pseudogap regime is probably the strangest one of all properties as it is a metallic phase with, as I will show, an energy gap that eliminates much of the Fermi surface. Although the phase diagram shown is for $La_{2-x}Sr_xCuO_4$ and $Nd_{2-x}Ce_xCuO_4$, two particular families of the curpates, the physical picture is more general. As indicated by the numbers at different places of the phase diagram, I intend to select seven examples to illustrate how ARPES has helped us gaining insight into the electronic structure and many-body effects responsible for the rich phase diagram. In this colloquial paper, I summarize only the experimental results from my own group and our collaborators presented at the conference. For more extensive reading on results and references from all groups and related theoretical ideas, the reader is referred to reviews and conference proceedings [6-12]. This paper is intended for a general reader, however I will make comments aiming at an expert reader of the high-$T_c$ field.

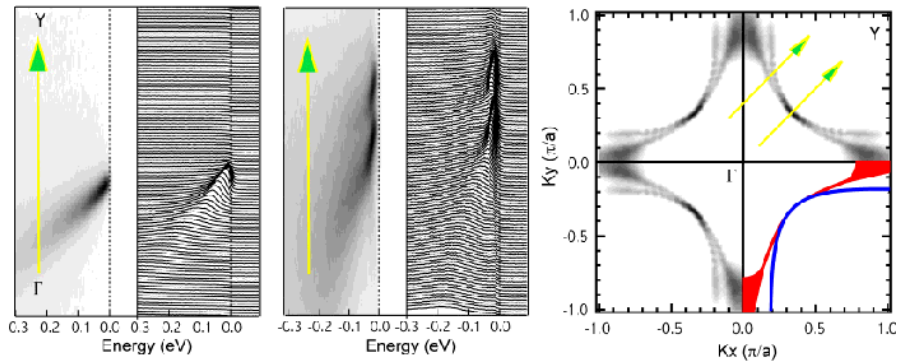

Figure 3, Fermi surface of overdoped $Bi_2Sr_2CaCu_2O_{8+\delta}$

The first example is a result from the deeply over-doped regime, where the physical properties are generally found to be less anomalous and conventional theory (Landau Fermi liquid theory) appears to be more applicable. The material chosen for the experiment is that of $Bi_2Sr_2CaCu_2O_{8+\delta}$ – a member of the Bi based cuprate family that yields a particularly stable surface. The right panel of the figure 3 displays the spectral intensity map at the Fermi level for the entire Brillouin zone [13]. Such a map normally gives a good representation for the Fermi surface – key microscopic information needed to calculate physical properties of metals. The other two panels show raw data from two sample-cuts in momentum space as indicated in the Brillouin zone. The energy versus momentum band dispersion is clearly discernable in the corresponding image plots. It is evident that there are two Fermi surface pieces in this material as marked by different colors. They stem from the fact that there are two $CuO_2$ planes in the unit cell, giving rise to bonding and antibonding combination of the Fermi surface. The detailed behavior of the splitting, zero along the (0,0)-(1,1) direction and growing away from it in a form of $[cos(k_xa)-cos(k_ya)]^2$, agrees well with predictions from band structure calculations, signals more conventional physics in this deeply overdoped case [14].

For an expert reader, it should be noted that the detection of this so-called bi-layer splitting of the Fermi surface is a testament to the precision of modern photoemission experiments, especially to the improvement in momentum resolution. The fact that this splitting was not detected in earlier experiments with poorer angular resolution was taken as evidence for electronic confinement within the planes and de-confinement driven superconductivity. The improved momentum resolution gave a different answer. Recent works by several groups have also indicated evidence of the splitting in optimal and slightly underdoped materials.

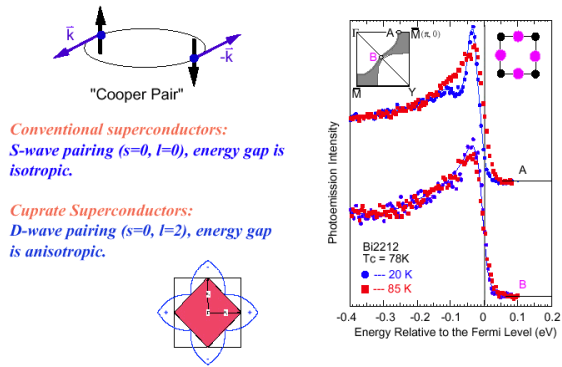

Figure 4. ARPES data supporting a d-wave superconducting gap structure. Left panel depicts the expected d-wave pairing state and its order parameter.

The second example is the detection of the d-wave gap structure as depicted in figure 4 [15]. Although we show data from an overdoped $Bi_2Sr_2CaCu_2O_{8+\delta}$ ($T_c$ = 78K), this result for the gap structure at low temperature is more generally valid. A key step for superconductivity is the formation of Cooper pairs where two electrons (fermions) bind to form a pair (boson). In conventional superconductors, the pairing is generally in the s-wave channel where the orbital angular momentum of the pair is $l = 0$. In this case, the superconducting gap, which is a measure of the binding energy of the pair, is isotropic. In the cuprate superconductors, consideration of strong Coulomb and/or magnetic interactions led to preferential pairing in the d-wave channel where $l = 2$. This d-wave pairing leads to a strong momentum space anisotropy in the superconducting gap that can be directly measured by ARPES. In particular, d-wave pairing dictates that the gap is zero along the zone diagonal direction, a consequence of a sign change of the orbital pair wave-function across this line. As shown in the right panel, the spectra taken along this direction (point B) do not show any change above and below $T_c$. In a striking contrast, the spectra taken along the bond direction (point A) show a remarkable difference above and below $T_c$. The shift of the leading edge unambiguously indicates the opening of a superconducting gap below $T_c$. This result and the result from the microwave measurement of the London penetration depth provided crucial early evidence for a $d_{x2-y2}$ pairing state in cuprates[2], and helped to stimulate the debate and motivate additional experiments that led to the current consensus on the pairing symmetry.

For the expert reader, the following points on this subject are worth noting. First, aside from the unconventional pairing symmetry, the superconducting state in cuprates is actually more conventional than the normal state. For example, the Meissner effect and the flux quantization are what one expects from a typical superconductor. The d-wave pairing symmetry is fully compatible with the Bardeen-Cooper-Schrieffer theory of superconductivity. Second, the behavior that a gap opens only below $T_c$ is a characteristic of deeply overdoped samples – an issue we will come back to. Third, the pairing symmetry alone cannot determine the pairing mechanism that can be more complicated. While d-wave pairing is likely an indication for the strong role of Coulomb interactions, and the quest to determine the pairing symmetry has been largely motivated by the magnetic pairing scenarios, it in principle does not rule out a lattice pairing mechanism. I note that the reason for a particular pairing symmetry does not have to be the reason to give the high $T_c$. It is possible that several factors may conspire to enhance $T_c$ and to yield the rich results observed in cuprates.

The third example is the electronic structure of undoped antiferromagnetic insulator. Figure 5 shows the band dispersion in

$Sr_2CuO_2Cl_2$ that has the same $CuO_2$ plane [16]. This material was chosen for its superior surface quality as compared to other cuprate insulators. The left and the middle panel show raw data and the experimental dispersion that is compared with band theory, respectively. In dramatic contrast to the overdoped case shown in figure 3, the agreement between the data and the band calculation is rather poor. While the theoretical dispersion shows a maximum at $(\pi,\pi)$, the experimental maximum is reached half way at $(\pi/2,\pi/2)$. This can naturally be explained by the zone folding effect as the antiferromagnetic order doubles the unit cell and thus reduces the Brillouin zone to half (see low right corner). Indeed, the experimentally determined dispersion can be well accounted for by theoretical models that takes into account the magnetic exchange interaction J, such as the so-called t-t'-t'-J model in figure 6. In this case, the hopping parameters (t, t', t'') are from the band structure calculation and the magnetic interaction energy J is independently determined by neutron scattering experiments. It is evident that the inclusion of magnetic interaction significantly improved the agreement with experiment – both in term of symmetry breaking and the magnitude of dispersion. This in tern suggests that the magnetic interaction is essential to understanding the physics of the cuprates, especially in undoped and underdoped cases. The strong contrast between the undoped case here and that of the deeply overdoped case in figure 3 mirrors the dramatic change of other physical properties as indicated in the phase diagram of figure 2.

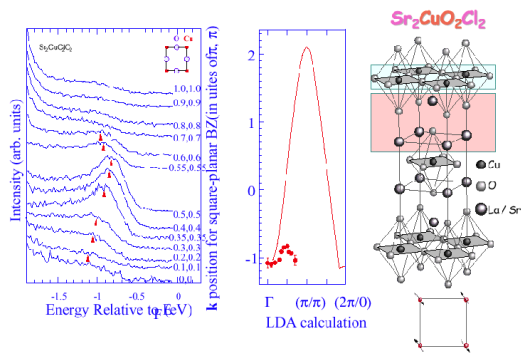

Figure 5. ARPES data from insulating $Sr_2CuO_2Cl_2$, whose structure is depicted in the right panel. The left panel shows the raw EDC along $(0,0)$-$(\pi,\pi)$ line. The middle panel compares the experimental dispersion with that of the band structure calculation.

For the expert reader, a few remarks are in order. First, an ARPES study of the insulator is in fact a measurement of the motion of a single hole (generated by the photoemission process) in the antiferromagnet. This problem turns out to be an excellent model system to test many-body theory. Despite its simplicity, exact solution of the t-J model (or the related Hubbard model which with appropriate parameters can also account for the dispersion in figure 6) is only possible numerically on small clusters. For realistic parameters, the solution for the single-hole case is the most robust one and the magnitude of the dispersion is determined only by the exchange interaction J. With J being independently determined by neutron scattering, this experiment provides a stringent test for theoretical calculations and thus has been very influential and widely cited. The physical reason for the dispersion being determined by J rather than the bare hopping t is that the hole motion disrupted the antiferromagnetic background which in turn slows it down. In a related measurement on a one-dimensional material [17], it was found that this slowing down of electron motion does not happen. This result can naturally be explained by the so-called spin-charge separation, a special behavior of one-dimensional systems, where the doped hole decays into a spinon and a holon whose motion is not influenced by the spin system. This experiment complements the example of the two-dimensional system in showing the importance of magnetic interactions. Secondary, not all aspects of the data (as well as the later data by many groups) can be explained by the t-t'-t''-J model, among them is the anomalous peak width of the spectrum at ($\pi/2,\pi/2$). The anomalously strong temperature dependence of the spectra cannot be accounted for either. These effects may be related to lattice effects in this material, as we will discuss later.

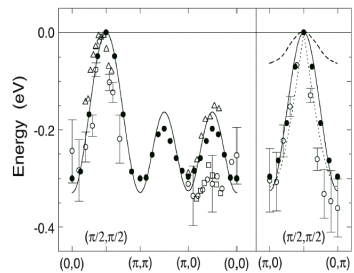

Figure 6.
Compiled ARPES data in comparison with theoretical results from t-t'-t''-J model.

The fourth example of important ARPES experiments on cuprates is that of the normal state gap in the underdoped regime, also referred as pseudogap. The original observation was made on the $Bi_2Sr_2CaCu_2O_{8+\delta}$ compound [18, 19]. We present here the simpler case of $Ca_{2-x}Na_xCuO_2Cl_2$ – a compound with only one $CuO_2$ sheet in the unit cell and thus simpler

Fermi surface [20, 21]. As shown in the left panel of figure 7, a large Fermi surface centered at the (π,π) point is expected. However, the experimental data only shows Fermi surface crossings near the dots of the middle panel, as if the Fermi surface terminates in the middle of Brillouin zone. As the area enclosed by the Fermi surface is a measure of the number of fermions, the Fermi surface must close and cannot terminate in this fashion. The most natural explanation is that given in the right panel, namely portion of the Fermi surface is gapped even at temperatures above the superconducting transition temperature. This gapping behavior is highly anomalous for a metal and is now referred to as one of the most telling aspects of the unusual normal state properties of the cuprate superconductors. In fact, this region of the phase diagram marked as the pseudogap phase is often taken as an important anchoring point for theoretical models. The pseudogap effect is indicated by many experimental techniques, the ARPES work has made a significant impact by virtue of its directness in revealing the momentum dependence. For slightly underdoped samples, the pseudogap can best be characterized by a shift of the spectral leading edge away from the Fermi energy (typically of the order of 25 meV). Detailed measurements reveal that both the magnitude and the angular dependence of the pseudogap have a striking similarity to that of the superconducting gap below $T_c$. This provides strong support for the most natural explanation of the pseudogap phenomenology, namely that the pair formation above $T_c$.

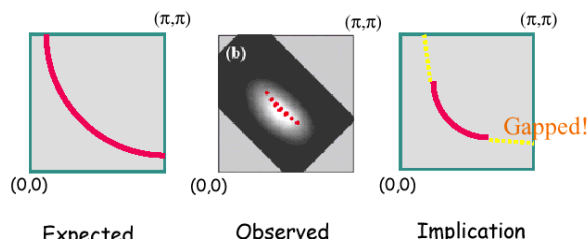

Figure 7. ARPES pseudogap phenomenology. Left panels depicts what is expected, middle panel shows the data and right panel indicates the gapping effect.

For an expert reader, it should be noted that there is another aspect of the pseudogap that cannot be explained by the pair formation above $T_c$. This is the so-called high-energy pseudogap that has an energy scale much larger than the one we discussed above that is of the same magnitude as the superconducting gap (sometimes referred as the low-energy pseudogap in the literature). The high-energy pseudogap refers to a suppression of the spectral density at an energy scale comparable to that of the magnetic interaction J. This aspect is most evident in deeply underdoped cases, and is apparently connected to the fact that the superconductor is smoothly connected to the antiferromagnetic insulator by doping. Indeed, the dispersion of the insulator in figure 6 allows such a connection. In this sense, for much of the underdoped regime, the data is a combination of the two effects – a leading edge shift of smaller energy scale in the **k**-space region where the d-wave gap is a maximum and a spectral weight suppression over a large energy scale. It is presently unclear whether the two effects are caused by the same underlying physics or not.

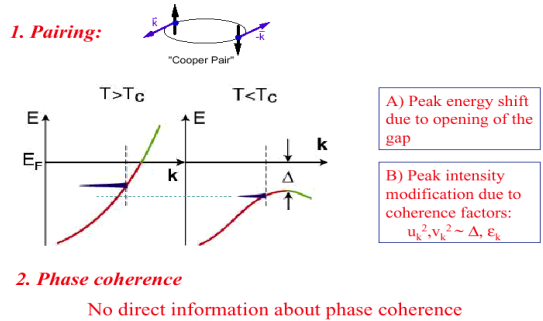

Figure 8. Cartoon depicting the expected ARPES result from the mean-field BCS theory.

Given prospect of gap (and pair) formation in the normal state, the fifth example we will discuss is the nature of the superconducting transition in the underdoped regime [22]. To set the stage, it is instructive to review the expected ARPES results from the BCS theory as illustrated in figure 8. In BCS theory, the superconducting transition is characterized by the opening of the superconducting gap that results in a *shift* of the quasiparticle peak energy and a *reduction* of the peak intensity due to coherence factors $u_\mathbf{k}^2$ and $v_\mathbf{k}^2$. ARPES measurement gives no direction information on phase coherence – a necessary condition for superconductivity that is implicitly assumed in a mean field theory like BCS. As shown in figure 9 for data from $Bi_2Sr_2CaCu_2O_{8+\delta}$, the ARPES result is completely different from the BCS theory. The superconducting transition is instead characterized by the emergence of a sharp peak whose intensity increases with lowering temperature. Note that the normal state spectra do not show a peak at all, and its leading edge energy position is pulled back from the Fermi level by about 25 meV, manifestation of the small pseudogap. There is no shift of the leading edge energy position across $T_c$. These results made plain that the superconducting transition is characterized not by the opening of an energy gap, but rather the emergence of a coherent state where sharp excitations are possible. Please note that the same behavior is also seen in the slightly overdoped sample, albeit that the superconducting gap is smaller and the low temperature peak is bigger. There is a gradual cross-over from one regime to another.

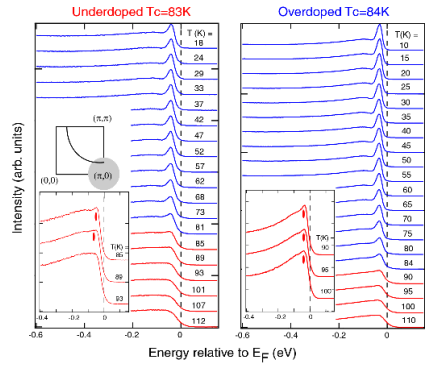

Figure 9.
ARPES spectra recorded near the (π,0) point from a lightly underdoped sample ($T_c$~ 83K) and a slightly overdoped sample ($T_c$~84K).

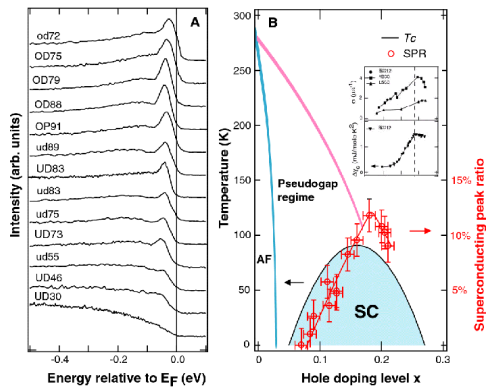

Figure 10.
Left panel, low temperature ARPES spectra near (π,0) as a function of doping. Right panel, the superconducting peak intensity (ratio of the peak again the broad background) as a function of doping. Inset, superfluid density from μSR and condensation energy.

The anomalous temperature dependence can further be corroborated by the doping dependence. The left panel of figure 10 shows the low temperature spectra from $Bi_2Sr_2CaCu_2O_{8+\delta}$ recorded at the **k**-space region where the superconducting gap is the maximum [(π,0) region of the Brillouin zone]. The right panel plots the peak intensity normalized by the background at higher energy as a function of doping. In the underdoped regime, the peak intensity monotonically increases with doping, following the rise of $T_c$. Curiously, the peak intensity reaches a maximum near 18-20%, above the optimal doping for $T_c$ near 15%. As shown in the inset, this is exactly what is also seen in the condensation energy measured by the specific heat and the superfluid density measured by muon spin resonance. As the later two quantities measure the superconducting condensate fraction, empirically the temperature and doping dependence established that the superconducting peak intensity also measures the superconducting condensate fraction, completely different from what is expected from the BCS theory. An understanding of this feature will lead to considerable insight into the nature of cuprate superconductivity.

For the expert reader, a few comments are worth noting. Indeed, the superconducting transition in the underdoped regime is likely governed by phase coherence among the pairs that exist at higher temperature, and the complete absence of the sharp peak above $T_c$ is likely a consequence of phase fluctuation. The fact that the disappearance of the sharp peak above $T_c$ is not due to a trivial thermal smearing is confirmed by the study of the $Bi_2Sr_2Ca2Cu_3O_{10+\delta}$ compound with a $T_c$ of 110K [23]. While the trivial thermal broadening is not very different from 85K, in this case the sharp peak appears abruptly below 110K. This study also found that the increased $T_c$ in this system stems from an increase in both the peak intensity and the size of the superconducting gap – indicating that the both an increase in condense fraction and the pairing strength. However, phase fluctuation alone does not explain why the low temperature peak intensity decreases with a doping decrease in a fashion that is exactly like the superfluid density. Fermi liquid like ideas to explain the decrease of the peak intensity by the increase of the pseudogap and the band position at $(\pi,0)$ are completely wrong, as such a coincidental scenario cannot explain the striking correlation seen in figures 9 and 10.

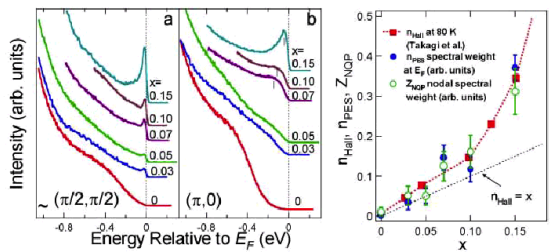

Figure 11. Data from $La_{2-x}Sr_xCuO_4$ near $(\pi/2,\pi/2)$ and $(\pi,0)$ as a function of doping. Right panel, peak intensity near $(\pi/2,\pi/2)$ shown together with the Hall numbers of the same system.

To drive home this point, we show in figure.11 ARPES data from the $La_{2-x}Sr_xCuO_4$ system where the quality of the crystal is better in the underdoped regime than for $Bi_2Sr_2CaCu_2O_{8+\delta}$, which is especially true for deeply underdoped cases [24]. For spectra taken near the Brillouin zone center $(\pi/2,\pi/2)$ where the d-wave gap is zero, we can see the same effect where the peak intensity decreases with the doping decrease in the same way as that of the Hall coefficient (right panel) and thus the superfluid density. Here, there is no gapping effect at all as the measurement is done along the nodal direction. We remark here that the band near $(\pi/2,\pi/2)$ is more dispersive so the feature can be more easily washed out by **k**-averaging from disorder. Underdoped $Bi_2Sr_2CaCu_2O_{8+\delta}$ is more disordered hence one cannot see this effect for the nodal quasiparticles when their spectral weight is small. As the spectra in figure 11 are recorded around 25K for $La_{2-x}Sr_xCuO_4$ where $T_c$ varies from 0K to 40K, it is interesting that one sees a sharp peak near $(\pi/2,\pi/2)$ even above $T_c$ for cases when $T_c$ is low but not near $(\pi,0)$ even below $T_c$ for the case when $T_c$ is high. This contrasts to the more disordered $Bi_2Sr_2CaCu_2O_{8+\delta}$ system where one does not see a sharp peak at $(\pi/2,\pi/2)$ at any temperature and only below $T_c$ at $(\pi,0)$ for cases where $T_c$ is above 40K. This strongly suggests a hierarchy of disorder effects. The material disorder is most disruptive at $(\pi/2,\pi/2)$ but not very disruptive at $(\pi,0)$ where the band is very flat, and this explains the $Bi_2Sr_2CaCu_2O_{8+\delta}$ situation. On the other hand, phase disorder more quickly destroys the peak at $(\pi,0)$ in the normal state where the gap is the largest. The fact that no sharp peak is seen in the cleaner $La_{2-x}Sr_xCuO_4$ at $(\pi,0)$ even below $T_c$ is likely due to the fact that $T_c$ of this system is very low, close to the lowest measuring temperature. Indeed, no underdoped $Bi_2Sr_2CaCu_2O_{8+\delta}$ sample with $T_c$ below 40K shows a sharp peak at $(\pi,0)$ either.

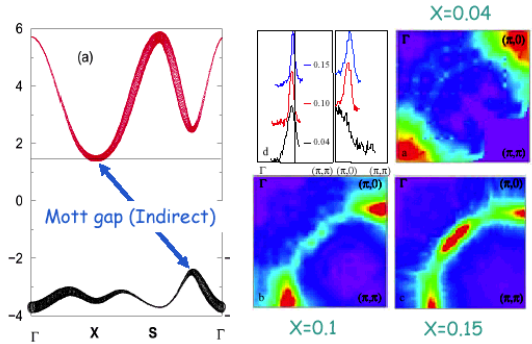

Figure 12. Left panel, a mean-field t-t'-t''-U model. Right panel, Spectral intensity at the Fermi level of x=0.04, 0.1, and 0.15 of $Nd_{2-x}Ce_xCuO_4$, respectively.

The sixth example where ARPES has given considerable insight on the cuprate problem is the evolution of the normal state electronic structure with doping, an issue also related to the so-called high-energy pseudogap. This issue is best addressed by the electron-doped $Nd_{2-x}Ce_xCuO_4$ system as in this case we can see the destruction of the Mott gap more clearly [25, 26]. The left panel of figure 12 depicts a Hubbard fit to the "valence band" data below the gap of the insulator (the data from insulating $Nd_2CuO_4$ is very similar to those in figure 5) and the expected "conduction band". This fit dictates that the Mott insulator is not a direct gap insulator. While the valence band maximum is at $(\pi/2, \pi/2)$ and is more relevant to the p-type doping, the conduction band minimum occurs near $(\pi,0)$, which makes this **k**-space region more relevant to electron doping. As shown in the near $E_F$ spectral weight map for the 4% doped sample, the Fermi surface is a small pocket near $(\pi,0)$. With the increase of electron doping, the measured Fermi surface evolves in a highly non-trivial fashion. At 10% doping, one sees that the $(\pi,0)$ Fermi surface pocket has changed its shape and at the same time some faint spectral weight start to emerge near the $(\pi/2,\pi/2)$ region. At 15% doping, the near $(\pi,0)$ Fermi surface feature evolves even more and a sharp Fermi surface arc appears near $(\pi/2,\pi/2)$ region. The connection of these two pieces gives a large circular Fermi surface centered at $(\pi,\pi)$, something very reminiscent to what is expected from band structure calculations. These results vividly demonstrate the distinction between doping a Mott insulator and a conventional band insulator. The evolution of the Fermi surface cannot be explained by a rigid filling of the "band structure" in the left panel, instead the underlying band structure evolves dynamically with doping.

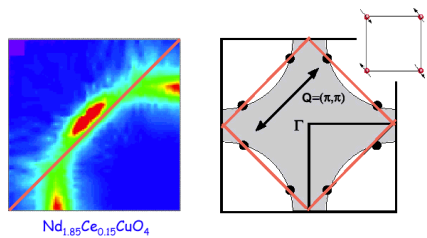

Figure 13. Left panel shows the Fermi surface intensity map from x=0.15 $Nd_{2-x}Ce_xCuO_4$ sample. Right panel depicts the Fermi surface hot spots intercepted by the antiferromagnetic zone boundary.

More insights on the dynamic evolution of the electronic structure can be gained by examining the Fermi surface of the 15% doped case as shown in figure 13. It is apparent that the Fermi surface feature is sharp and well defined except the regions that are intercepted by the antiferromagnetic Brillouin zone boundary as depicted by the diagonal line. Examination of the energy spectra along the Fermi surface verifies that the energy spectrum is sharp along the Fermi surface except these intercepted regions where the spectrum is very broad. This picture is what is expected in theoretical scenarios where the Umklapp scattering across the antiferromagnetic zone boundary ($\mathbf{Q}=(\pi,\pi)$) creates the so-called "hot spots" that truncate the underlying Fermi surface. This again suggests that the magnetic interaction is important to understanding the cuprate physics. Thus, one may also try to understand the evolution of the electronic structure starting from the metallic side, just opposite to the sequence of our discussion in figure 12. With the increasing strength of the antiferromagnetic interaction upon decreasing doping, the Fermi surface is truncated and the band is further folded and the Fermi surface evolves into small pockets when the long-range magnetic order sets in below 13% doping. In either case, the magnetic interaction (or alternatively the Coulomb interaction) plays an essential role for the electronic structure evolution.

For the expert reader, the following issues are worth noting. First, the band structure in figure 12 stems from a mean-field approach to the Hubbard model, and may be used as a cartoon to guide the discussion on a few salient features in the electronic structure that appear to be more generally true. However, dynamics are essential to understand the actual experimental spectra. A detailed discussion of these issues is beyond the scope of this paper. Second, the suppressed spectral weight near $E_F$ around the "hot spots" verifies the second aspect of the pseudogap phenomena, namely the suppression the spectral weight over an extended energy range. This effect is also present in the p-type cuprates, but it can be most unambiguously identified in the n-type case because the interception between AF zone boundary and the underlying Fermi surface occurs away from the ($\pi$,0) region so that this effect is distinct from that of a d-wave like gap. The stronger antiferromagnetic interaction in the n-type case may also play a role here. The truncation of the Fermi surface discussed in figure 13 and the identification of truncated Fermi surface pieces near the ($\pi$,0) region with n-type carrier and that near the ($\pi/2,\pi/2$) region with p-type carriers may explain many unusual normal state properties. For example, transport data suggest the presence of both p- and n-type carriers – a puzzle that may be explained along the line of the Fermi surface truncation.

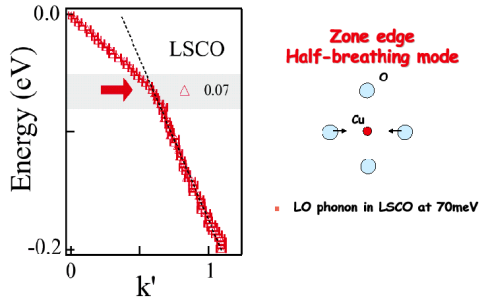

Figure 14. ARPES derived dispersion from 10% doped $La_{2-x}Sr_xCuO_4$ system. A sudden change of dispersion is seen. The red arrow (and the gray bar) illustrates the energy where in plane phonon (as shown in the right panel) softening is observed.

Having extensively discussed the physics involving charge and spin degrees of the freedom, we now discuss the issue concerning the lattice degree of freedom. Our last example of ARPES investigation of the many-body physics in cuprates has to do with the evidence for strong electron-phonon interaction. Given the fact that electron-phonon interaction is the mechanism for conventional superconductivity, this problem is of considerable importance. The role of electron-phonon interaction in cuprates is very controversial. An important reason for this is the lack of direct evidence of lattice effects on the electron self-energy as routinely seen in the classical tunneling experiments on conventional superconductors. In its most simple form appropriate for this article, the ARPES manifestation of an electron's interaction with phonons (or any collective mode for that matter) is often a sharp kink in the electron's energy-momentum dispersion near the phonon energy that is accompanied by an abrupt drop in the electron's scattering rate at the same energy. Figure 14 shows the dispersion of 10% doped $La_{2-x}Sr_xCuO_4$ near the Fermi surface crossing along the (0,0) – ($\pi,\pi$) direction. It is evident that the dispersion shows a sudden change near 70 meV, exactly where the in plane oxygen bond stretching mode showed anomalous softening as found by neutron scattering experiments (indicated by the red arrow) [27]. The striking co-incidence between the anomalous electronic and phononic self-energy effects at the same energy makes a likely case for strong electron-phonon coupling. As the interaction with the lattice has changed the electron's velocity (which is equal to the slope of the dispersion curve) by 100% above and below the phonon energy, we believe that the lattice effect is essential to the cuprate physics. The ARPES evidence for strong electron-lattice interaction is again a testament to the improvement in resolution, especially the momentum resolution. Until confronted by the striking data such as those in figure 14 (and figure 15 below), it was hard to predict whether ARPES can be an effective tool to address this issue. Indeed, this advance was made mostly experimental and empirical with much prior input from theory.

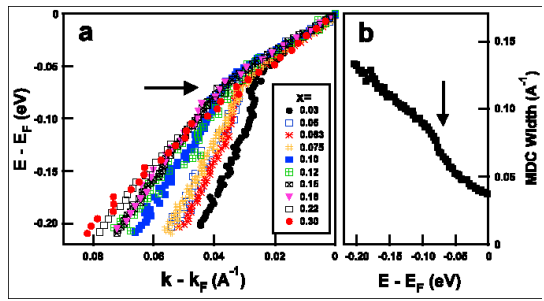

Figure 15. Left panel, dispersion of LSCO from x = 0.03 to x=0.3. Right panel, scattering rate (reflected in the width of the so-called momentum distribution curve) for x=0.63 sample.

For the expert reader, several comments are needed. Although the sudden dispersion change near 70 meV results from the oxygen lattice effect, the doping dependence reveals that the electron-lattice interaction in cuprates is highly anomalous. Figure 15 shows the doping dependence of the dispersion for $La_{2-x}Sr_xCuO_4$ system along the $(0,0)$–$(\pi,\pi)$ direction [28]. Interestingly, the velocity of these nodal electrons, i.e. the slope of the dispersion, remains the same within 70 meV of the Fermi energy over a wide doping range going from underdoped non-superconductors, underdoped superconductors, the optimal and overdoped superconductors, as well as overdoped non-superconducors. Such a universal velocity is in strong contrast to the rapid change of many other physical properties with doping. For the velocity at energies more than 70 meV away from the Fermi level, the data show an anomalous *increase* in velocity with *decreasing* doping. This gives an increased effective electron-lattice coupling strength with decreasing doping. However, it should be stressed that this occurs in a way that is completely different from the simple Fermi liquid picture where one expects that the velocity beyond the phonon energy from the Fermi level ("bare velocity") to be universal while the velocity within the phonon energy from the Fermi level ("renormalized velocity") to vary with change of electron-phonon coupling strength, just the opposite to what is observed. It is beyond the scope of this paper to discuss the implications of the behavior seen in figure 15, it is sufficient to note here that the anomalous velocity change beyond the phonon frequency may be related to the poor screening of the long range Coulomb exchange interaction, an effect that gets enhanced with underdoping. This interaction is well screened under the plasmon frequency in typical metals. In cuprates, however, it is not well screened especially along the c-axis. In this sense, the optical phonon may play the role the plasmons do to screen the long range Coulomb interaction, and making the velocity below its frequency universal. The second issue to be noted is the possible interplay between the lattice effect and the superconducting energy gap, a possibility in systems with higher $T_c$ where the gap is bigger. Two aspects of this possibility deserve attention. The first has to do with the possibility of enhanced superconductivity when the two energy scales are near resonance while the second has to do with additional self-energy effects due to the opening of the superconducting gap. Indeed there is an additional kink in the dispersion, which is most prominently seen away from the $(0,0)$-$(\pi,\pi)$ direction and turns on only below $T_c$, and occurs at comparable energy scales. A detailed discussion of these issues is beyond the scope of this manuscript.

The above seven examples include only some of the important issues in cuprates. An important issue not included in our discussion is local inhomogeneity or nano-scale phase separation. In some cases, this takes the form of one-dimensional stripes. Several attempts have been made to address these issues, but the manifestation of these effects in ARPES is often quite subtle so I did not include them in this more general article. These issues are discussed in a more extended review [6].

Through the above seven examples, I hope to give the reader a snapshot of ARPES results from different regimes of the cuprate phase diagram. These results provide a microscopic foundation to understand the rapidly varying physical properties through the rich electronic structure changes. There are two aspects to these results. On one hand, they provide the basic band structure information that has its roots in the crystal structure and chemical bonds that form the material. On the other hand, they reveal important many-body physics involving charge, spin and lattice degrees of freedom, as exemplified by the folding and renormalization of the band structure as well as Fermi surface truncation at hot spots of the antiferromagnetic interaction (which has its root in the Coulomb interaction), the anisotropic superconducting gap and pseudogap, the coherence transition at $T_c$ as well as the electron-lattice interaction. As the important goal of solid state physics is to construct simple models, which although have their roots in the underlying lattice structure and chemical bonds, catches the central physics in question. The fact that ARPES is sensitive to both aspects makes it an ideal tool to study such many-body physics.

Looking towards the future, the field will continue to develop at the breathtaking pace as we have seen over the last decade. There will be a continued push for even better resolution in both energy and momentum, and application to an even wider range of problems. The materials will also improve considerably, and systems with extremely low dimension for which this technique has unique advantages will grow in importance as some of the most interesting physical phenomena occur in such systems. At the same time, concerted efforts will be made to perform high-resolution angle-resolved photoemission experiments with spin detection, and to perform experiments with very high spatial resolution that approaches the molecular level. These improvements will open new frontiers. We are witnessing a renaissance and transformation of a technique from that of chemical bonds to that of many-body physics, and can be sure of new discoveries and surprises.

References:

1. G.B. Levi, Physics Today **43** (3), 20 (1990).
2. G.B. Levi, Physics Today **46** (5), 17, (1993).
3. G.B. Levi, Physics Today **49** (1), 19, (1996).
4. S. Mitton, Science Watch **6** (1) 1, (1995).
5. S. Mitton, Science Watch **9** (2), 1, (1998).
6. A. Damascelli, Z. Hussain, Z.-X. Shen, Rev. Mod. Phys., **75**, 473 (2003).
7. M. Randeria and J.C. Campuzano, cond-mat/9709107.
8. D.W. Lynch and C.G. Olson, Photoemission Studies of High-Temperature Superconductors (Cambridge University, Cambridge) (1999).

9. M.S.Golden, C. Durr, A. Koitzsch, S. Legner, Z. Hu, S. Borisenko, M. Knupfer and J. Fink, J. Electron Spectrosc. Relat. Phenom. **117-118**, 203 (2001).
10. P.D. Johnson, A.V. Fedorov and T. Valla, J. Electron Spectrosc. Relat. Phenom. **117-118**, 153 (2001).
11. Proceedings of the International Conference on Materials and Mechnisms of Superconductivity, High-Temperature Superconductors V, Beijing, China. Edited by Yu-Sheng He, Pei-Heng Wu, Li-Fang Xu and Zhong-Xian Zhao, North-Holland (1997).
12. Proceedings of the International Conference on Materials and Mechnisms of Superconductivity, High-Temperature Superconductors VI, Houston, USA. Edited by Kamel Salama, Wei-Kan Chu and Paul C.W. Chu, North-Holland (2000).
13. P. Bogdanov, A. Lanzara, X.J. Zhou, W.L. Yang, E. Eisaki, Z. Hussain, Z.-X. Shen; Phys. Rev. Lett. 89, 167002 (2002)
14. D.L. Feng, N.P. Armitage, D.H. Lu, A. Damascelli, J.P. Hu, P. Bogdanov, A. Lanzara, F. Ronning, K.M. Shen, H. Eisaki, C. Kim, J.-i. Shimoyama, K. Kishio, and Z.-X. Shen; Phys. Rev. Lett. 86, 5550 (2001)
15. Z.-X. Shen, D.S. Dessau, B.O. Wells, D.M. King, W.E. Spicer, A.J. Arko, D.S. Marshall, L.W. Lombardo, A. Kapitulnik, P. Dickinson, S. Doniach, J. Dicarlo, T. Loeser, and C.H. Park; Phys. Rev. Lett., 70, 1553 (1993)
16. B.O. Wells, Z.-X. Shen, A. Matsuura, D.M. King, M.A. Kastner, M. Greven, and R.J. Birgeneau; Phys. Rev. Lett. 74, 964 (1995)
17. C. Kim, Z.-X. Shen, N. Motoyama, H. Eisaki, S. Uchida, T. Tohyama, and S. Maekawa; Phys. Rev. B56, 15589-95 (1997)
18. D.S. Marshall, D.S. Dessau, A.G. Loeser, C.H. Park, A.Y. Matsuura, J. Eckstein, I. Bozovic, P. Fournier, A. Kapitulnik, W.E. Spicer and Z.-X. Shen; Phys. Rev. Lett. 76, 4841 (1996)
19. A.G. Loeser, Z.-X. Shen, D.S. Dessau, D.S. Marshall, C.H. Park, P. Fournier and A. Kapitulnik; Science, Vol. 273, 325 (1996)
20. Y. Kohsaka, T. Sasagawa, F. Ronning, T. Yoshida, C. Kim, T. Hanaguri, M. Azuma, M. Takano, Z.-X. Shen, H. Takagi; Journal of Physical Society, Japan., submitted.
21. F. Ronning, T. Sasagawa, Y. Kohsaka, K.M. Shen, A. Damascelli, C. Kim, T. Yoshida, N.P. Armitage, D.H. Lu, D.L. Feng, L.L. Miller, H. Takagi, and Z.-X. Shen; Phys. Rev. B., accepted for publication.

22. D.L. Feng, D.H. Lu, K.M. Shen, C. Kim, H. Eisaki, A. Damascelli, R. Yoshizaki, J.-i. Shimoyama, K. Kishio, G. Gu, S. Oh, A. Andrus, J.O' Donnell, J.N. Eckstein, and Z.-X. Shen; Science, 289, 277 (2000).
23. D.L. Feng, A. Damascelli, K.M. Shen, N. Motoyama, D.H. Lu, H. Eisaki, K. Shimizu, J.-i. Shimoyama, K. Kishio, N. Kaneko, M. Greven, G.D. Gu, Z.J. Zhou, C. Kim, F. Ronning, N.P. Armitage, Z.-X. Shen; Phys. Rev. Lett. 88, 107001 (2002)
24. T. Yoshida, X.J. Zhou, T. Sasagawa, W.L. Yang, P.V. Bogdanov, A. Lanzara, Z. Hussain, T. Mizokawa, A. Fujimori, H. Eisaki, Z.-X. Shen, T. Kakeshita, S. Uchida; cond-mat/0209339.
25. N.P. Armitage, D.H. Lu, C. Kim, A. Damascelli, K.M. Shen, F. Ronning, Y. Onose, Y. Taguchi, Y. Tokura, and Z.-X. Shen; Phys. Rev. Lett., 87, 147003 (2001)
26. N.P. Armitage, F. Ronning, D.H. Lu, C. Kim, A. Damascelli, K.M. Shen, D.L. Feng, H. Eisaki, Z.-X. Shen, P.K. Mang, N. Kaneko, M. Greven, Y. Onose, Y. Taguchi, Y. Tokura; Phys. Rev. Lett., 88, 257001 (2002)
27. A. Lanzara, P.V. Bogdanov, X.J. Zhou, S.A. Kellar, D.L. Feng, E.D. Lu, T. Yoshida, H. Eisaki, A. Fujimori, K. Kishio, J.-I. Shimoyama, T. Nodall, S. Uchida, Z. Hussain, Z.-X. Shen; Nature 412, 510 (2001)
28. X.J. Zhou, T. Yoshida, A. Lanzara, P.V. Bogdanov, S.A. Kellar, K.M. Shen, W.L. Yang, F. Ronning, T. Sasagawa, T. Kakeshita, T. Noda, H. Eisaki, S. Uchida, C.T. Lin, F. Zhou, J.W. Xiong, W.X. Ti, Z.X. Zhao, A. Fujimori, Z. Hussain, and Z.-X. Shen; Nature, in press.
29. The data in the inset of the figure 10 come from the groups of Uemura (μSR) and Loram (specific heat). The cartoon in the right panel of figure 12 comes from the group of Markiewicz.

Acknowledgement: I thank the conference organizers for the opportunity to participate in the meeting honoring professor C.N. Yang's contribution to physics. Professor Yang is childhood hero who has been an inspirational force for me to become a professional physicist. I thank Peter Armitage and Felix Baumburger for proof reading this manuscript. Over the last decade, the Stanford ARPES program has been supported by the Materials Science Division of the DOE's Office of Basic Energy Sciences, the Division of Materials Research of NSF and the Office of Naval Research. The ARPES experiments have been performed at the